\documentclass[a4paper,fleqn,usenatbib]{mnras}

\usepackage{txfonts}

\usepackage[T1]{fontenc}
\usepackage{ae,aecompl}

\usepackage{graphicx}	

\title[Mass Evaporation of Globular Clusters]{Mass Evaporation Rate of Globular Clusters in a Strong Tidal Field}

\author[J. P. Madrid et al.]{
Juan P. Madrid,$^{1,2}$
Nathan W. C. Leigh,$^{3}$
Jarrod R. Hurley,$^{2}$
Mirek Giersz$^{4}$
\\
$^{1}$CSIRO, Astronomy and Space Science, PO BOX 76, Epping NSW 1710, Australia\\
$^{2}$Centre for Astrophysics and Supercomputing, Swinburne
University of Technology, Hawthorn, VIC 3122, Australia\\
$^{3}$Department of Astrophysics, American Museum of Natural History, Central Park West and 79th Street, NY 10024, USA\\
$^{4}${\it Nicolaus Copernicus Astronomical Centre, Polish Academy of Sciences, ul. Bartycka 18, PL-00-716 Warsaw, Poland}
}

\date{Accepted XXX. Received YYY; in original form ZZZ}

\pubyear{2017}

\begin{document}
\label{firstpage}
\pagerange{\pageref{firstpage}--\pageref{lastpage}}
\maketitle

\begin{abstract}

The  mass evaporation  rate  of  globular  clusters evolving  in  a  strong
Galactic  tidal  field  is  derived  through the  analysis  of  large,
multi-mass $N$-body simulations.  For comparison, we also 
study the same evaporation rates using MOCCA Monte Carlo models 
for globular cluster evolution.  Our results show that the mass evaporation 
rate is a dynamical value; that is, far from a constant single number 
found in earlier analytical work and commonly  used in the  literature. 
Moreover, the evaporation rate derived with  these simulations  
is higher than values previously published.   These models also  
show that  the  value of  the mass evaporation rate  depends on  the strength of  
the tidal  field.  We give  an analytical estimate  of the
mass evaporation  rate as a  function of  time and  galactocentric distance
$\xi(R_{GC},t)$. Upon extrapolating this formula to smaller $R_{GC}$ 
values, our results provide tentative evidence for a very high $\xi$ value 
at small $R_{GC}$. Our results suggest that the corresponding mass loss in the inner Galactic potential
could be high and it should be accounted for when star clusters pass within it.  This has direct 
relevance to nuclear cluster formation/growth via the infall of globular clusters 
through dynamical friction.  As an illustrative example, we estimate  how 
the evaporation rate increases for a $\sim 10^5$ M$_{\odot}$ globular cluster 
that decays through dynamical friction into the galactic centre.   
We  discuss the  findings  of  this work  in relation  to the  formation 
of  nuclear star  clusters  by inspiraling globular clusters.

\end{abstract}

\begin{keywords}
Globular Clusters: general -- Stars: kinematics and dynamics -- Methods: numerical
\end{keywords}



\section{Introduction}

A  more  detailed examination  of  the  evaporation  rate of  globular
clusters ($\xi$)  obtained through  modern $N$-body models  is granted
given that the  value of $\xi$ is often used. A  recent example is the
work  of \citet{gnedin2014} that  uses the  evaporation rate  in their
study  of nuclear star  cluster formation  through the  inspiraling of
globular clusters  into the galactic nuclei.  The  evaporation rate is
also   used  by  \citet{gieles2011}  in their  study  of  the
evolution of star clusters in a tidal field.

\citet{ambartsumian1938}    and   \citet{spitzer1940}    established   a
dimensionless  evaporation   rate  $\xi$  that   takes  the  following
expression:

\begin{equation}
\xi \equiv - \frac{t_{rh}}{M}\frac{dM}{dt}
\label{eqn:dimensionless}
\end{equation}

where $t_{rh}$ is the half-mass relaxation time given by

\begin{equation}
t_{rh}= \frac{0.14N}{\ln \Lambda} \sqrt{\frac{r_{hm}^3}{GM}},
\label{eqn:trh}
\end{equation}

where $\Lambda =  0.4N$ is the argument of  the Coulomb logarithm, $N$
is  the  number of  stars,  $M$ the  globular  cluster  mass, $G$  the
gravitational   constant,   and    $r_{hm}$   the   half-mass   radius
\citep{spitzerhart1971}. We should note that for single mass models $M$
is proportional to  $N$. 

Initial   estimates   of   $\xi$   were   obtained   analytically   by
\citet{ambartsumian1938}  who   found  $\xi  =0.0074$   and  later  by
\citet{henon1961}        who         found        $\xi        =0.045$.
\citet{spitzerchevalier1973} recalculate  the value of  $\xi$ and find
this quantity is  dependent on the ratio between  the tidal radius and
the  half-mass  radius  $r_{t}/r_{hm}$.   \citet{spitzerchevalier1973}
find $\xi$= 0.015 and $\xi$=0.05 for different $r_{t}/r_{hm}$ ratios.

\citet{leegoodman1995} carried out a detailed study of the evaporation
rate   for   postcollapse  globular   clusters   using  a   multi-mass
Fokker-Planck  code.   They find  that  the  mass-loss  rate $\xi$  is
roughly constant for most of  the evolution of the globular cluster --
their Figure 1.

While the  publications cited above  are among those  that established
the  foundations   of  star   cluster  dynamics  their   results  were
nonetheless   obtained  making   several  approximations.   Among  the
simplifications  used are star  clusters in  isolation, that  is, free
from tidal  interactions, star  clusters with single-mass  stars, just
schematic  stellar evolution,  and star  clusters in  the  steady post
core-collapse regime.

The reality underlying cluster evaporation and mass loss, however, is much 
more complicated than these pioneering works cited above fully addressed.
For example, the inclusion of realistic stellar evolution and the associated 
mass loss breaks the one-to-one connection between mass loss and evaporation, since 
stars can reduce their masses due to stellar evolution without actually 
evaporating from the cluster.

We  have now  the  ability to  accurately  simulate globular  clusters
throughout  an  entire   Hubble  time  of  evolution.   \cite{bmakino}
performed a  series of $N$-body  models of globular clusters  in tidal
fields  using {\tt  NBODY4}.   These  models were  used  to show  that
stellar evolution  plays an  important role in  the mass-loss  rate of
globular  clusters,  particularly during  the  early  stages of  their
lifetimes.  \cite{bmakino} also  showed how  the external  tidal field
plays a crucial role in  determining the dissolution time for globular
clusters.   Other  studies   have  addressed  different  environmental
effects on  cluster disruption  and mass-loss such  as bulge  and disc
shocking      \citep{gnedin1997}       and      radial      anisotropy
\citep{brockamp,webb,webb2,leigh2013}.

\citet{hong2013} make a distinction between the evaporation rate of single 
and multimass models of globular clusters, i.e.\ $\xi(N) \neq \xi(M)$. 
\citet{hong2013} use a definition of the {\it mass evaporation rate} $\xi(M)$ 
that we adopt here. The mass evaporation rate that we refer through this paper 
is not limited to evaporation of stars from two body interactions. 
Our dimensionless mass evaporation rate includes all mechanisms of mass loss.
Our method folds into a single $\xi$ parameter all physical processes of mass
loss such as tidal interactions with the host galaxy, 
stellar evolution, relaxation driven mass loss, etc.

\cite{gieles2008}  studied  the  escape  rate of  stars  from  tidally
limited  star clusters  with  different radii.   The main  differences
between  this work  and the  work  of \cite{gieles2008}  are that  our
simulations include stellar evolution,  primordial binaries and a more
advanced  description of  the tidal  field. \cite{gieles2008} assumed a 
point mass potential. Also,  for multimass models, star-loss rates 
and mass-loss rates are not equivalent, we show this in the following sections.

Here, we study the dimensionless mass-loss rates using a more sophisticated 
treatment for the Galactic potential than adopted in previous studies, 
with an emphasis on the inner regions of the Galaxy. In our simulations 
the galactic potential is modelled with three distinct components: bulge, 
disc, and halo. The additive gravitational force of each of these three 
components is computed for each star of the simulation.


In this work  we derive the value of $\xi$ for  globular clusters in a
strong  tidal field using mainly $N$-body simulations, but also  use the 
Monte Carlo code MOCCA, which provides results with 
minimal computational expense, as well as a detailed stellar evolution 
prescription.  

Throughout this  work we  consider a  Hubble time to  be 13.8  Gyr, in
agreement with the latest cosmological findings of the \citet{planck2015}.


 \begin{figure*}
 \includegraphics[width=\textwidth]{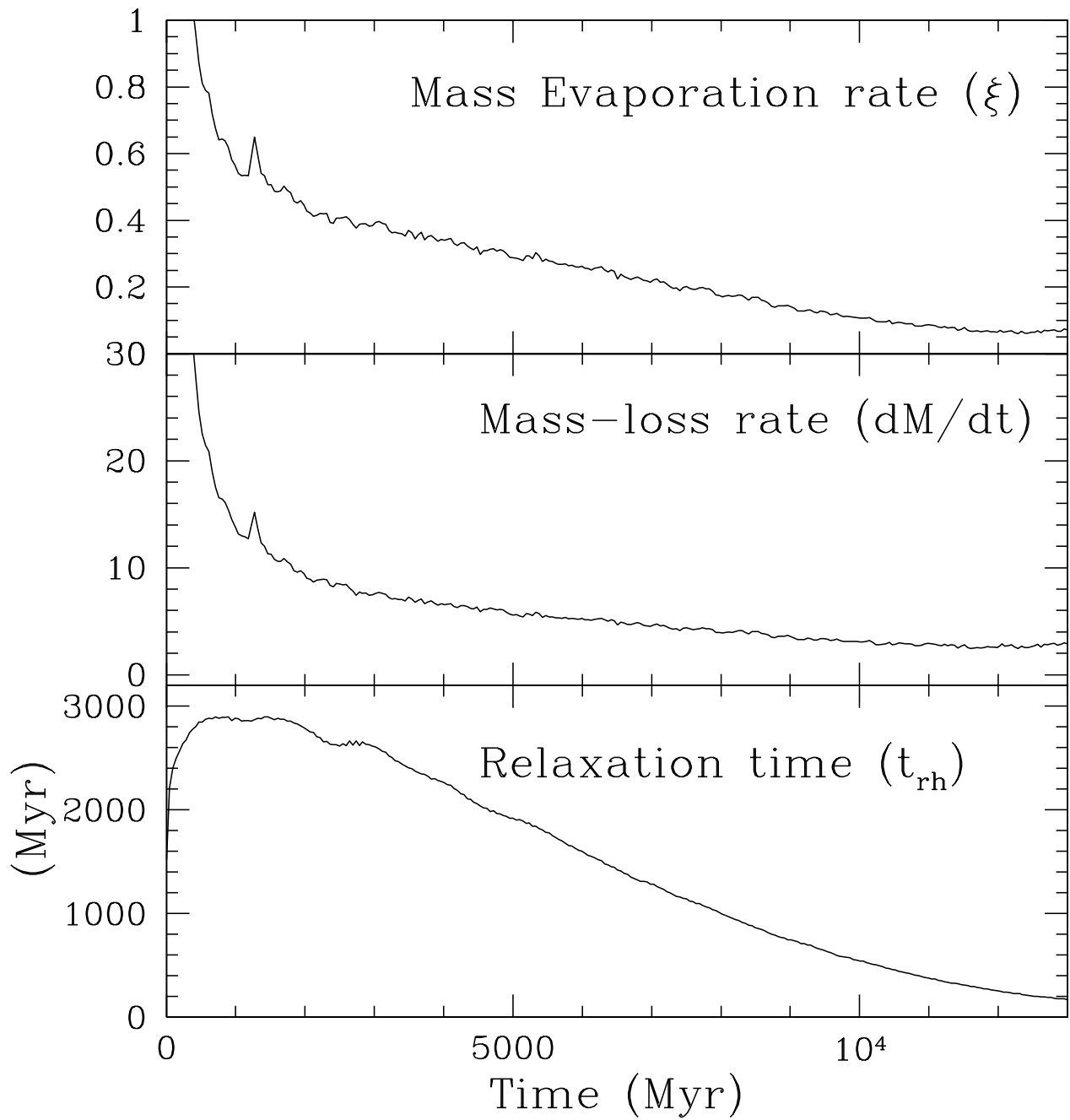}
 \caption{Mass evaporation  rate (top  panel) and  the two  physical values
   used  in Equation 1:  mass-loss rate  (middle panel)  and half-mass
   relaxation time  (bottom panel).  The  mass evaporation rate $\xi$  of a
   globular  cluster in  a strong  tidal  field varies  over time:  an
   initial rapid  decrease due to  stellar evolution is followed  by a
   linear decay phase. This figure uses the average values from models 
   1 \& 2 (Table 1).
 \label{fig1}}
 \end{figure*}


\section{A Simulated Globular Cluster Evolving in a Strong Tidal Field}

In this section, we describe the codes used to simulate globular cluster 
evolution.  We begin with the $N$-body simulations before moving on 
to describe the Monte Carlo models, since our focus in this paper are 
the former.  The Monte Carlo models are meant to complement and compare 
the results of our $N$-body simulations.

\subsection{$N$-body Simulations} \label{nbody}

\subsubsection{Models setup}

An up-to-date  version of the code  {\tt NBODY6} is used  to carry out
the   simulations  \citep{aarseth1999}.    This  version
includes the  gravitational potential of a Milky-Way  type galaxy \citep{aarseth2003}. 
The Galactic  potential is modelled  using a  bulge mass  of $M_B=1.5\times
10^{10}  M_{\odot}$ and  disc mass  $M_D=5 \times  10^{10} M_{\odot}$.
The  geometry  of  the  disc  is modelled  following  the  formulae  of
\citet{miyamoto1975} with  the following scale  parameters $a$=4.0 kpc
and  $b$=0.5  kpc.  A  logarithmic  halo  is  included such  that  the
circular velocity is 200 km/s at 8.0 kpc from the Galactic center. The
tidal field generated by the host galaxy is taken into account for the
calculation  of  the equations  of  motion  of  all stars  within  the
simulated globular cluster. Our simulations begin when all the stars in 
the globular cluster are formed, all residual gas removed and the cluster
is in a stable orbit around its host galaxy. The initial velocity distribution
of stars follows virial equilibrium.

The  simulations   were  run  on  Graphic   Processing  Units  (GPUs),
specifically NVIDIA Tesla C2070 cards mounted on the GPU Supercomputer
for Theoretical Astrophysics Research at Swinburne University (gSTAR).
The set-up of these models  has previously been described in detail in
\citet{madrid2012} and \citet{madrid2014}.

We define a escape radius for stars as it was previously defined in 
\citet{madrid2012} and \citet{madrid2014}. That is, stars are lost from the 
simulation when they have positive energy and when their distance to the
center of the cluster exceeds two tidal radii. The tidal radius \citep{vonhoerner}
is a dynamical quantity, which is in turn we define using the approximation 
of \citet{kupper10}:

\begin{equation}
r_{\rm t} \simeq \left( \frac{GM_{\rm C}}{2\Omega^2} \right)^{1/3}.
\label{eqn:eqkk}
\end{equation}

where $\Omega$ is the angular velocity.


\subsubsection{Initial Conditions}

In  this  section  we describe  a  series  of  $N$-body models  with  a
large  number of  star-particles, i.e.\ $N$=  200~000 (Labels 1 \& 2, Table 1). Five
per cent of these stars are in  binary systems, that is, a total of ten
thousand stars  are binaries. These simulations are  among the largest
direct   N-body   simulations  carried   out   to   the  present   day
\citep{heggie2011, heggie2014, wang2016}.  Each model takes about two 
months to be completed running on one GPU. Simulations with  even  larger 
numbers  of particles and  initial masses would need much more computing 
time, since this time-scale is proportional to $N^{\sim2-3}$, where $N$ is 
the number of particles.

These  simulations  are multi-mass  $N$-body  models  that follow  the
prescriptions  of the  \citet{kroupa2001} initial  mass function
(IMF).   This  IMF  and  $N   =200~000$  yields  an  initial  mass  of
$M=1.3\times 10^5 M_{\odot}$. The minimum stellar mass is $M_{min}=0.1 M_{\odot}$ 
and the maximum stellar mass is  $M_{min}=50 M_{\odot}$. NBODY6 includes stellar 
and binary evolution \citep{hurley2000,hurley2002}. 

The  initial spatial distribution of the globular cluster  follows 
a \citet{plummer1911} model  and the initial metallicity  is $Z=0.001$.  
We should note that the Plummer model is an idealised spatial distribution 
altered during the first evolutionary steps by the host galaxy tidal field 
and internal dynamical evolution. The simulated  globular cluster  analyzed 
in this  section was  set to evolve  at  a galactocentric  distance  of  
$R_{gc}=4$  kpc.  At  this distance  from  the galactic  centre  the 
simulated globular  cluster experiences a  strong tidal field which is 
the additive gravitational potential  of bulge and  disc \citep{madrid2014}.

The initial half-mass radius is $r_{hm}=6.2$ pc, a slightly extended cluster,
and the initial tidal radius according to the standard formula is $\sim 52$ pc,
with the cluster tidally-filling at birth. Stars are allowed to have orbits that
go beyond the tidal radius and then return to the star cluster. As mentioned before, 
stars are only removed from the simulation when their orbits exceeds more than two tidal radii. 

As formulated in \citet{madrid2014}, our simulated star clusters have an initial 
inclination of $\sim~22.5$ degrees with respect of the disc. With this inclination 
star clusters are simulated in a general orbit as opposed to a specific 
co-planar or perpendicular inclination with respect of the disc. The initial inclination
we use also implies that the star cluster has a maximum height of two kiloparsecs which
is similar to the disc scale height.

In the  interest of reducing numerical  uncertainties simulations were
carried out twice  (Labels 1 \& 2, Table 1) with a different seed number  for the random number
generator used in the  distributions that describe the initial masses,
positions  and  velocities of  the  stars.  Results  of the  otherwise
identical simulations are then averaged.


\subsection{Monte Carlo Simulations} \label{mocca}

\subsubsection{Models setup}

An up-to-date  version of the {\tt MOCCA} code is used  to carry out
the Monte Carlo simulations for globular cluster evolution  \citep{giersz13,hypki13}. 
The  gravitational potential of the Milky-Way is modeled with a much simpler point-mass, 
with a total mass equal to that of the enclosed Galaxy mass at the cluster's Galactocentric 
distance $R_{GC}$ from the Galactic Center.  We should note that at equal host galaxy mass, 
and equal galactocentric distance the tidal field is stronger in a point mass potential 
than with a logarithmic potential. All orbits through the Galactic  potential are circular 
within the MOCCA framework.

The simulations were performed on a PSK2 cluster at
the Nicolaus Copernicus Astronomical Centre in Poland. Each
simulation is run on one CPU. The cluster is based on AMD
(Advanced Micro Devices, Inc.) Opteron processors with 64-bit
architecture (2-2.4 GHz).  The set-up of these models  has previously been 
described in detail in \citet{giersz13} and \citet{leigh2013}.

The MOCCA code treats the escape process in tidally limited clusters in a 
realistic manner as it is described in \citet{fukushige2000}.
Here, the escape of an object from the system is not instantaneous, but
delayed in time. As was pointed out in \citet{fukushige2000} and
\citet{baumgardt2001}, the process of escape from a cluster in a steady tidal
field is complex. Some stars that fulfill the energy criterion
for escape (i.e.\ the condition that the energy of the star exceeds the
critical energy) can still be trapped inside the potential well. Some of
those stars can be scattered back to lower energies before they escape
from the system. These two factors cause the cluster lifetime to scale
non-linearly with relaxation time for tidally limited clusters \citet{baumgardt2001},
in contrast to what would be expected from the standard theory,
see details in \citet{giersz13}. The MOCCA simulation results are presented 
in Figure~\ref{fig5}. To minimize the statistical fluctuations connected with generation 
of the MOCCA initial models (labels 9 \& 10, Table 1) we used exactly the same masses, 
positions and velocities as for the $N$-body runs (labels 1 \& 2, Table 1).


\begin{table}
	\centering
	\caption{Parameters of Simulated Star Clusters.}
	\label{tab:example_table}
	\begin{tabular}{lcccr} 
		\hline
		Label & $R_{GC}$ & $N$ & $M_i$ & Code\\
		\hline
		1 & 4 kpc & 200k & 1.3$\times 10^5$ M$_{\odot}$ & NBODY6\\
		2 & 4 kpc & 200k & 1.3$\times 10^5$ M$_{\odot}$ & NBODY6\\		
		3 & 4 kpc & 100k & 6.3$\times 10^4$ M$_{\odot}$ & NBODY6\\
        4 & 6 kpc & 100k & 6.3$\times 10^4$ M$_{\odot}$ & NBODY6\\
        5 & 8 kpc & 100k & 6.3$\times 10^4$ M$_{\odot}$ & NBODY6\\
        6 & 10 kpc & 100k & 6.3$\times 10^4$ M$_{\odot}$ & NBODY6\\  
        7 & 20 kpc & 100k & 6.3$\times 10^4$ M$_{\odot}$ & NBODY6\\     
        8 & 50 kpc & 100k & 6.3$\times 10^4$ M$_{\odot}$ & NBODY6\\   
        9 & 4 kpc & 200k & 1.3$\times 10^{5}$ M$_{\odot}$ & MOCCA\\
        10 & 50 kpc & 100k & 6.3$\times 10^{4}$ M$_{\odot}$ & MOCCA\\       				
		\hline
	\end{tabular}
\end{table}



\section{Mass Evaporation Rate}

Previous  studies  cited above  consider  the  evaporation  rate as  a
constant  value during  the  entire lifetime of a  globular cluster.   As
pointed out by \citet{takahashi2000} the value of the evaporation rate
in the  calculations of \citet{leegoodman1995}  is always less  than 1
($\xi < 1$) due to the  absence of stellar evolution. In the models of
\citet{leegoodman1995} the  evaporation rate increases  with time from
$\sim  0.002$ at  the beginning  of the  globular cluster  lifetime to
$\sim  0.05$. After  the cluster  has  evolved for  $\sim$30\% of  its
lifetime  the  evaporation   rate  derived  by  \citet{leegoodman1995}
asymptotes to a constant value  of $\sim 0.05$ until the cluster fully
dissolves.   The results  of the  $N$-body and MOCCA models  performed  for this
study show  that the  behaviour of the  mass evaporation rate for  a cluster
under  a  substantial tidal  field  varies  over  time and  cannot  be
approximated by a constant single value,  as shown in the top panel of
Figure 1.  As mentioned above in the introduction, our dimensionless mass
evaporation rate folds into a single parameter all mechanisms of mass loss.

When stellar  evolution is included in   the mass evaporation rate, 
as we do here, $\xi$ reaches its highest values at  the beginning
of the globular  cluster lifetime and then  decreases, which  is the
opposite  of the  results presented  by \citet{leegoodman1995}.

The initial value of the mass evaporation rate for
the  models represented in  Figure 1  is $\xi  = 6.9$.   The mass-loss
induced by stellar  evolution rapidly tapers off within  the first Gyr
of the  star cluster  lifetime and the  evaporation rate  decreases to
$\xi < 1$ before 500 Myr.

The mass evaporation rate reaches its  lowest value of $\xi~=~0.07$ after a
Hubble time  of evolution, see Fig.\  1 top panel. Even  at its lowest
value  $\xi$  is  higher   than  some  previous  analytical  estimates
\citep{henon1961,leegoodman1995}.  The  much higher value  of $\xi$ is
due  to  the  realistic  modelling of  the  tidal  field and  the
inclusion of stellar evolution in the simulations.

The  mass evaporation  rate plotted  in  Figure  1  reflects the  different
dominant regimes of mass-loss  a globular cluster undergoes: mass-loss
due to  stellar evolution, and  mass-loss owing to  tidal interactions
and  internal  dynamics  \citep{lamers2010,kruijssen2012,leigh2012b}. 

\citet{lamers2010} find three different phases of mass loss for star 
clusters: (A) mass loss dominated by stellar evolution, (B) mass loss dominated by 
``dissolution" or dynamical effects, (C) and the third phase being after core 
collapse. Our models are in good agreement with the results of \citet{lamers2010}. 
In particular, we find good agreement on the rapid and early mass loss due to stellar 
evolution. We also agree on the second phase of mass loss being dominated 
by tidal interactions and dynamical processes. This second phase is fitted with
a power law by \citet{lamers2010} similar to what we describe and fit as 
a linear decay.
A difference with the work of  \citet{lamers2010} is that the simulated 
globular  clusters discussed in this  section do  not  reach core  collapse 
during a  Hubble time  of evolution. Core  collapse is  actually reached 
at  15 Gyr.

The  mass evaporation rate  and  the  first time  derivative  of the  total
globular cluster  mass ($dM/dt$)  have a bump  at $\sim$1.3  Gyr. This
bump is  due to stellar evolution  processes.  At 1.3 Gyr  is when the
main-sequence  turn-off   mass  is   $\sim  1.8$  M$_{\odot}$   for  a
metallicity  of $Z  =  0.001$ \citep{pols1998}.   This is  significant
because  it is  the   mass  at  which   stars  switch   from  having
non-degenerate  to  degenerate  cores  on  the  giant  branch.   Thus,
post-giant  branch timescales  change and  the stellar  mass-loss rate
changes. Also, for any binary interactions that strip giant envelopes,
the resulting star  will now be a helium white  dwarf instead of going
through a  naked helium star phase,  which also affects  the amount of
mass-loss \citep{hurley2000, hurley2002}.

The evolution of the half-mass relaxation time closely follows that of
the half-mass  radius. For example, the early  increase in $t_{rh}$ is
the result  of an  initial expansion phase  for the cluster. Stellar 
evolution is the main driver of mass loss during the early lifetime of
globular clusters. Rapid mass loss causes,  in turn, a reduced gravitational
pull resulting in an expansion. For tidally filling clusters, after the initial expansion, 
the  half-mass radius  steadily decreases as  the tidal radius  of 
the  cluster  decreases   and  in  turn  we  see  $t_{rh}$ decreasing. 
A detailed study of the size scale of star clusters is given in 
\citet{madrid2012}.

NBODY6 properly accounts for the important effects of disc shocking when the 
star cluster is formed and settled in a stable orbit \citep{madrid2014}. 
However, the early effects of impulsive shocks in the gas-rich progenitor 
galaxy \citep{kruijssen2015} or the interactions of the star cluster with giant 
molecular clouds are not modelled in the current version of NBODY6. More 
sophisticated hydrodynamic simulations are needed to properly parameterize 
this correction to the dimensionless mass loss rate.

What is the dependence of the mass evaporation rate on the initial mass of
the star cluster? We can briefly address this question by comparing the two models 
with different initial masses ($1.3\times10^5$ M$_{\odot}$ and $6.3\times10^4$ M$_{\odot}$
-- labels 1 and 3 (Table 1). For a star cluster under the influence of a strong tidal field,
i.e. on an orbit with $R_{GC}=4$kpc, the mass evaporation rate is almost identical
during the early phases of evolution when $\xi$ is dominated by mass loss 
due to stellar evolution. The overall evolution of $\xi$ is equivalent  for the 
two models with different masses
that we consider.



\subsection{Total mass and number of stars}


 \begin{figure}
 \includegraphics[width=\columnwidth]{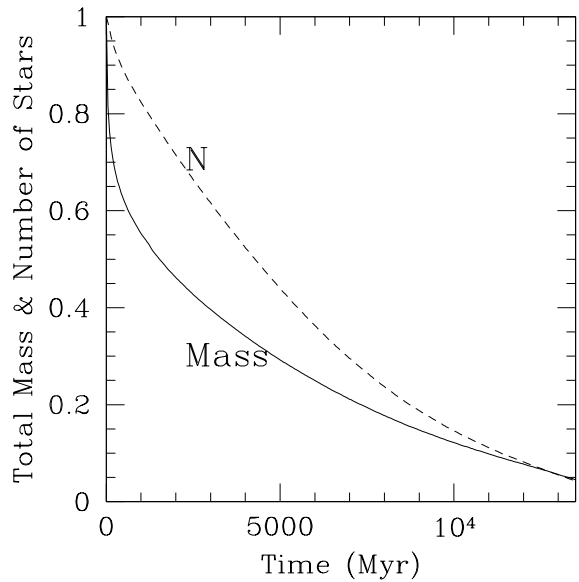}
 \caption{Total mass  and number of  stars remaining in  the simulated
   star  cluster.   These quantities  have  been  normalized by  their
   initial values. This Figure represents the average values for models 1 \& 2 on Table 1.
 \label{fig2}}
 \end{figure}


The total mass  and number of stars that remain  bound to the globular
cluster are  plotted in  Figure 3. Mass-loss  rates of  large $N$-body
simulations  have been  discussed  elsewhere \citep{madrid2012}.   The
inclusion of stellar evolution makes the globular cluster lose mass at
a faster rate than the rate at which stars are lost during the initial
12 Gyr of evolution. The mass-loss rate and star loss rate osculate at
12 Gyr and are almost identical until dissolution.
 
The  behaviour  of   the  mass  and  number  of   stars  over  time  is
fundamentally different  from that obtained with early  work with very
low $N$.   In the star clusters  simulated by \citet{gierszheggie1996}
mass-loss and  star-loss rates are identical during  the initial phases of
evolution to later continuously diverge until dissolution. This difference
can be explained, at least in part, by a different IMF and the fact that 
the maximum stellar mass is higher in our simulations.


 \begin{figure}
 \includegraphics[width=\columnwidth]{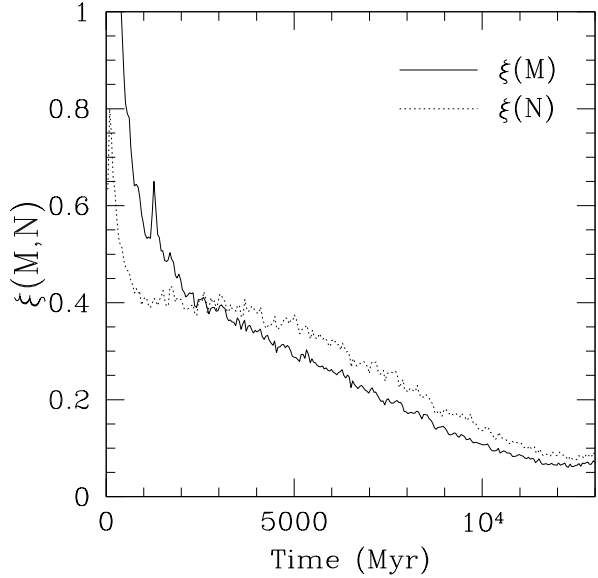}
 \caption{Dimensionless evaporation rate as function of mass $\xi(M)$ and 
as a function of number of stars $\xi(N)$. $\xi(N)$ is obtained by simply
replacing $M$ by $N$ in Eq.\ 1. The difference between the two curves reflects
the fact  these are multimass models that include a prescription for 
stellar evolution. These are the average values for models 1 \& 2 on Table 1.
 \label{fig3}}
 \end{figure}


Also,  the evolution of  the total  number of  stars ($N$)  over time,
plotted in Figure 2, is curved  and different from the linear relation 
followed by $N$ in the models of \citet{gieles2011}. Our results suggest 
higher star-loss rates than those of the models presented by \citet{gieles2011}.

As mentioned in previous sections, earlier analytical work mostly used
single mass models where $\xi(M)  \equiv \xi(N)$ \citep{gieles2011}.  In Figure 3 we plot
the value  of $\xi$ as  a function  of mass (M)  and as a  function of
number of stars  that remain bound to the cluster  (N).  The latter is
obtained replacing  M by N on  Eq.\ 1. The difference  between the two
expressions  is naturally  more  pronounced during  earlier stages  of
evolution ($ < 3$ Gyr), when $\xi(N)$ is less than $\xi(M)$.  The main
factor contributing  to this difference  is stellar evolution  and the
fact that we ran multimass models. Figure 3 shows that at about 4 Gyr of 
tidal striping takes over stellar evolution as the main mass-loss mechanism.


\section{Recovering Earlier Analytical Findings}

In this section we compare the results for a globular cluster evolving
in a strong tidal field and a globular cluster in pseudo-isolation.
We  also show  that the  $N$-body  and MOCCA models presented  here can  recover
earlier analytical findings discussed in Section 1.

Figure 4 shows  the time evolution of $\xi$  for two globular clusters
in  different  tidal  environments, taken from the results of our $N$-body 
simulations.    We  plot  in  this  figure  the
evaporation rates  for globular clusters  on circular orbits at  4 kpc
(same  as  Fig.\  1)  and  50  kpc  from  the  Galactic  center.   The
dimensionless mass-loss rate is  clearly different and thus a function
of the tidal field.

The   mass evaporation   rate  of   the   globular   cluster  evolving   in
pseudo-isolation at  50 kpc (label 8, Table 1) from  the galactic centre asymptotes  to a
nearly constant value  after $\sim$4 Gyr  of evolution.  This behaviour  is in
good agreement with previous  analytical work that considered globular
clusters evolving in isolation. The numerical value of the evaporation
rate  derived with our  simulations at  $R_{GC}=50$ kpc  asymptotes to
$\xi   \approx   0.04$  thus   recovering   earlier  analytical   work
\citep{henon1961}.

\section{Comparison with MOCCA Monte Carlo models}

For comparison, we show in Figure 5 the time evolution of the mass evaporation rate
for two globular clusters once again in different tidal environments, 
taken from the results of our MOCCA Monte Carlo models.  These models 
are chosen to have identical initial conditions as the 
$N$-body models shown in Figure 4 (see Section 2.2). We carefully select
MOCCA models that are tidally filling at small $R_{GC}$ and tidally 
underfilling at large $R_{GC}$. Two MOCCA models at Galactocentric 
radii of 50 kpc (black dots), and 4 kpc (blue squares) kpc are shown in Fig.\ 5.

Figure 5 shows that the MOCCA simulations are in good
agreement with the mass loss rates $\xi$ seen in the $N$-body 
models. In particular, MOCCA and $N$-body models agree well on the elevated 
mass loss rates due to stellar evolution seen up to about 3 Gyr. Later on,
the MOCCA model at $R_{GC}= 4$ kpc evolves faster than  the $N$-body model 
counterpart, that is, this MOCCA model reaches core-collapse faster than 
the $N$-body model. Post core-collapse, MOCCA models recover earlier analytical 
predictions for $\xi$ cited in the Introduction and section above.
The slightly greater value of $\xi$ for the MOCCA model evolving at $R_{GC}= 4$ kpc
can be explained by the stronger tidal field generated by the point mass potential
at small galactocentric distances.

We also see increased scatter in the MC results, which is due to the fact that, 
when calculating $\xi$ at each time-step, the MC method must sample randomly 
along the stars' orbits to obtain their positions, which induces artificial 
fluctuations in $\xi$ between time-steps.  The $N$-body models avoid this 
issue altogether by tracking the positions of each star directly, at each time-step.


 \begin{figure}
 \includegraphics[width=\columnwidth]{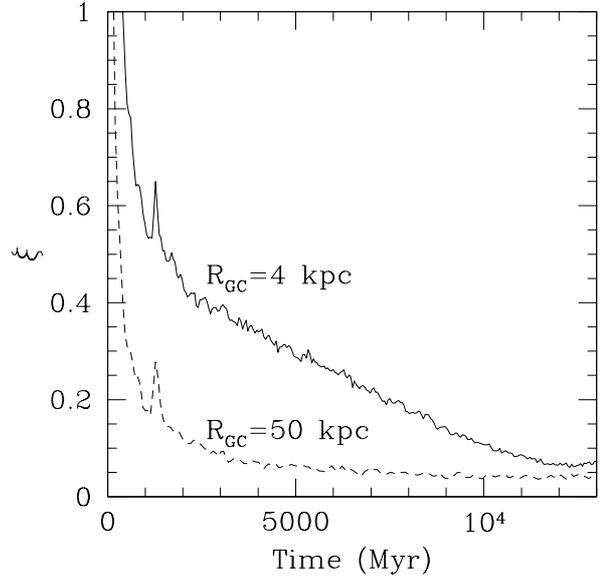}
 \caption{Time  evolution of  the dimensionless  mass evaporation rate in our 
 $N$-body models for a
   globular cluster  in pseudo-isolation  ($R_{GC}$ = 50  kpc, dashed)
   and a globular  cluster in a strong tidal field  ($R_{GC}$ = 4 kpc,
   solid). From the average values of models 1 \& 2 (at $R_{GC}$ =  4 kpc) and 
   model 8 (at  $R_{GC}$ = 50 kpc).
 \label{fig4}}
 \end{figure}

 \begin{figure}
 \includegraphics[width=\columnwidth]{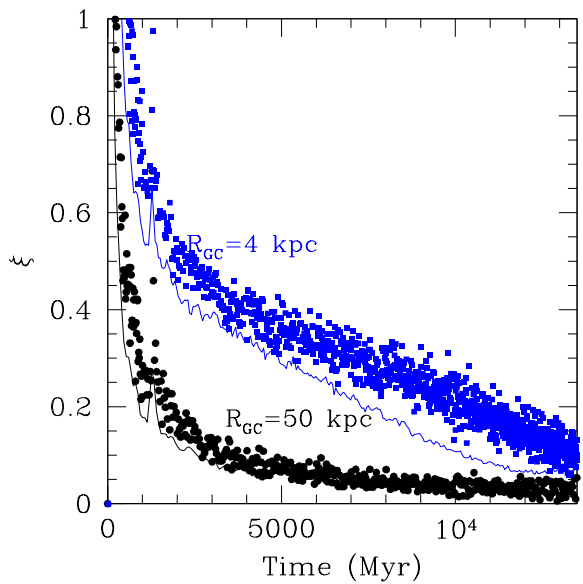}
 \caption{Time  evolution of  the dimensionless  mass evaporation rate in our MOCCA 
 Monte Carlo models for a globular cluster  in pseudo-isolation  ($R_{GC}$ = 50  kpc, 
 black dots) and a globular  cluster in a strong tidal field ($R_{GC}$ = 4 kpc,
   blue squares). The results of the N-body models shown in Fig 4 are also
   plotted here as solid lines.
 \label{fig5}}
 \end{figure}


\section{Evaporation Rate as a Function of Time and Galactocentric Distance}

It can be seen in Figures 1, 3, and 4 that the time evolution of $\xi$
between 3 and  10 Gyr can be approximated by a  linear function of the
simplest form:  $g(t)=a\times t +b$.   This linear phase  of evolution
for $\xi$ follows  the rapid decline from large  values due to stellar
evolution.

We fitted linear functions of the  form described above to a series of
$N$-body  models  of  globular  clusters at  different  galactocentric
distances in  order to make  a sketch of  
the evolution of $\xi$  as a function  of  time  and  as  a function  
of  R$_{GC}$,  thus  deriving $\xi(R_{GC},t)$.

The models we  use to derive $\xi(R_{GC},t)$ are  the models presented
in  \citet{madrid2012}. These  are models  with an  initial  number of
stars $N$=100  000 and  initial mass  $M_{init}\approx  6.3 \times
10^{4} M_{\odot}$. We  use the models that evolved at 4,  6, 8, 10, 20
and 50 kpc from the galactic centre (labels 3 through 8, Table 1). 
We have taken the approach of keeping the initial half-mass radius identical for
each model. This means the models at small galactocentric
distance start tidally-filling and become progressively  
tidally under filling as $R_{GC}$ increases.

We obtain the  following expression for $\xi(R_{GC},t)$

\begin{equation}
\xi(R_{GC},t)= a\times\log(t) + b
\end{equation}

where $a(R_{GC})$ and $b(R_{GC})$ are

\begin{equation}
a(R_{GC})=-10^{(3.0\times \exp(-4.0\times \log (R_{GC} -1.05)) -1.2)}
\end{equation}

and 

\begin{equation}
b(R_{GC})=10^{(2.5\times \exp(-4.0\times (\log (R_{GC} -0.4)) -0.5)}
\end{equation}

with $t$ in Myr and $R_{GC}$ in kpc, we give the details of its derivation in the appendix.



In Figure  6 we  use Eq.\  4 to sketch  the evolution  of $\xi$  for a
globular cluster that decays into the galactic centre. 
We consider a globular cluster that inspirals into the galactic 
centre through dynamical friction. The dynamical friction timescale is given 
by \citet{binneytremaine1987} in their equation (7-26). A globular cluster  
with an initial  mass of  $M_{init} \sim  10^5  M_{\odot}$, and  orbiting 
at $R_{GC}=15$ kpc  would take $\sim$13000 Gyr to decay into the galactic 
centre by dynamical friction alone. That is, a thousand times a Hubble time. 
The mass evaporation rate and the distance to the centre of the galaxy are depicted 
on the top panel of Figure 6.  $\xi$ undergoes a marked increase  as the globular 
cluster approaches the center of the galaxy (see below). We do not directly address
the magnitude of the dynamical friction time-scale, as this has been looked at 
in detail in previous studies e.g., \citet{gnedin2014}.

For a globular cluster that inspirals to the galactic centre 
the  dimensionless mass evaporation rate, as given by  Eq.\ 4, increases by a factor of 32
during the  time the orbit of the globular cluster decays from
$R_{GC} = 14$ kpc to $R_{GC} =  4$ kpc. The trend is clear towards the
centre of the galaxy where our results suggest that $\xi$ continues to
increase  within the inner  4 kpc.   As $\xi$ is extrapolated inwards, 
it re-approaches the very high values seen very early on due to stellar evolution.  
We should note that the factor of 32 increase is also dependent on the initial
conditions of the star cluster, for instance its initial half-mass radius.

The extrapolation of  our results to lower galactocentric distances  
illustrates that the value of $\xi$ could reach or even exceed the previous 
maximum value of $\sim$6.9 due to stellar evolution.  This result serves to highlight 
the importance of developing a better understanding on how a globular cluster responds
to  extreme tidal striping by the galactic potential in the inner regions of a galaxy. 
A more  detailed analysis  of the behaviour of  $\xi$ towards  the core of  the galaxy, 
including  a more detailed modelling of the bulge potential, will be presented by Rossi et
al.\ (in preparation).

 \begin{figure}
 \includegraphics[width=\columnwidth]{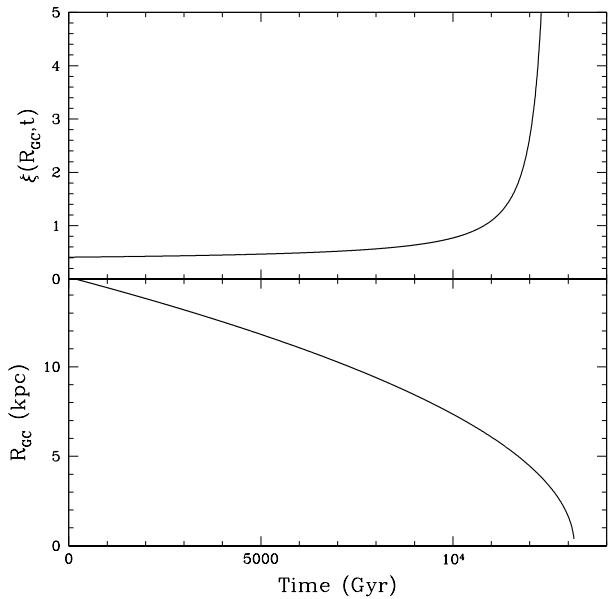}
 \caption{Evolution of  $\xi(R_{gc},t)$  for a  globular cluster that 
 decays  into the galactic centre. The top panel depicts the evolution 
 of $\xi$ for a globular cluster  that inspirals into the galactic 
  centre following the  dynamical friction timescale given by Binney 
  and Tremaine (1987). We find a marked increase of the value of  $\xi$ as the 
  globular cluster approaches the centre of the galaxy. Note that top and
  bottom quantities scale as the inverse of the other, this is highlighted by
  the log $y$-scale of the bottom panel.
 \label{fig6}}
 \end{figure}

Our results support the conclusions reached by \citet{webb15}, 
namely that modern simulations of GC evolution are still under-estimating the mass loss.  
The authors presented a method for estimating the total initial masses in Galactic 
globular clusters, using only the total cluster luminosity or mass and the slope of the 
present-day stellar mass function. The results of  \citet{webb15} are in good agreement 
with earlier findings of \citet{vesperini} and \citet{kruijssen2009}.

An  evaporation  rate  that  clearly  rises as  the  globular  cluster
approaches  the galactic  center  should certainly be accounted for in future 
models that propose to explain the formation of nuclear star clusters by 
infalling globular clusters e.g., \citet{gnedin2014}.

In terms of the overall mass loss rates, using the recent results of \citet{webb15}
and  \citet{cai2016}, we expect that the time-averaged mass loss rate for any given 
eccentricity can be approximately set equal to the rate for a circular orbit at a given 
Galactocentric distance, which is slightly larger than pericentre, typically, 
but obviously smaller than apocentre for the eccentric orbit.  It is not known, 
however, how an eccentric orbit affects the dynamical friction time-scale itself. 
Functions like the ones provided for $\xi(R_{GC},t)$ can be used to properly calculate 
the fate of a GC that loses mass as it spirals into the nucleus due to dynamical friction.


\section{Formation of Nuclear Star Clusters by Inspiraling Globular Clusters}

Globular clusters  are affected by dynamical friction  that makes them
inspiral  into the  center  of their  host  galaxy, possibly  providing
material   to  create   nuclear  star   clusters  \citep{tremaine1975,
  leigh2012,antonini2013,gnedin2014, arca2014}.

A caveat to the above theory is the dissolution of infalling globular
clusters before  reaching the nuclear region.  Dynamical  friction is a
function of  mass. While  a globular cluster  as an entire  entity can
experience  enough  dynamical friction  to  infall  into the  galactic
centre, its individual  stars, once  lost  due to  tidal heating,  are
practically unaffected by dynamical friction \citep{tremaine1975}.

Dynamical  friction  timescales  dictate  that only  massive  globular
clusters  that orbit  a galaxy  within a  few ($\sim  3-6$  kpc) would
spiral to  the galactic center during  a Hubble time,  see the example
for the  Andromeda galaxy by \citet{binneytremaine1987}.  We have shown
here that it is precisely in  this region that the evaporation rate of
globular clusters  is higher than previously thought.   This implies a
smaller  mass budget  to build  nuclear star  clusters  from infalling
globular clusters. Our models show that any globular cluster spiraling
into the nucleus will lose a significant fraction of its total initial mass, 
with that fraction being close to unity \citep{webb15}.

\citet{arca2014} recently modelled the  origin of nuclear star clusters
from migratory globular clusters.  They  found that when the effect of
tidal  shocks were  included in  their models  the number  of globular
clusters  that  effectively spiral  into  the  nucleus  of the  galaxy
decreases by a  factor of five.  \citet{arca2014} even  found that for
their  models of  massive galaxies  with tidal  heating the  number of
globular  clusters falling  into the  galactic centre  drops  to zero,
preventing the formation of a nuclear star cluster.

Through this  work we have shown  that when stellar  evolution and the
galactic tidal field are included into the models of globular clusters
its  dimensionless mass evaporation  rate  is much  higher than  previously
presented.   We   have  also  derived  the  dependence   of  $\xi$  on
galactocentric distance and time  quantifying how the evaporation rate
of globular clusters is enhanced as they approach the galactic centre.

We would like to highlight the importance 
of properly understanding mass loss in the inner Galactic potential, which 
is still highly uncertain.  In particular, our results show that mass loss 
in the inner Galaxy could be significant.  Even accounting for {\it all}
sources of mass loss in our estimate for $\xi$, our theoretical extrapolation
suggests it could rise above even the $\xi$ values we see at early times 
when stellar evolution dominates (included in our $\xi(M)$ estimate).
Although our simple theoretical extrapolation of the mass loss rate 
to the inner Galactic potential diverges at $r=0$ (see Figure 5), 
the true mass loss rate should not become infinite.  Whether or not the true 
mass loss rate does indeed get this high is an open question, and could 
significantly affect previous estimates of the mass loss rates used for 
calculations involving nuclear star cluster formation via globular cluster
inspiral.


\section*{Acknowledgements}

We would like to thank the anonymous referees for providing detailed reports 
that helped us improve this paper. This research  has made use of the  NASA 
Astrophysics Data System  Bibliographic  services  (ADS)  and  Google.   
This  work  was performed on  the gSTAR national  facility at Swinburne  
University of Technology.   gSTAR   is  funded  by  Swinburne   and  the  
Australian Government Education Investment Fund.   This research was supported by
Australian Research Council funding through grant DP110103509. MG acknowledges
support from the National Science Centre through the grant 
DEC-2012/07/B/ST9/04412.




\appendix

\section{Derivation of $\xi (R_{GC},t)$}

This appendix presents the details  of the derivation of $\xi (R_{GC},
t)$. In order  to derive $\xi (R_{GC},t)$ we  assume that $\xi$ varies
smoothly as  a function  of $R_{GC}$ and  that it takes  the following
general functional form:

\begin{equation}
\xi(R_{GC}, t)=f(R_{GC})\times g(t)
\end{equation}

for some unknown functions $f$ and $g$.

We also  assume that the  temporal component takes the  following form
$g(t)=a\times log(t) +b$.  As mentioned  in Section 5, the linear part
of the temporal evolution of $\xi$ for the numerical models is between
3 and 10  Gyr. Replacing $g(t)$ into the  general expression for $\xi$
we obtain:

\begin{equation}
\xi(R_{GC}, t)=f(R_{GC})\times (a\times \log(t) +b)
\end{equation}

or,

\begin{equation}
\xi(R_{GC}, t)=f(R_{GC})\times a\times \log(t) +f(R_{GC})\times b
\end{equation}

By absorbing the constants $a$ and  $b$ into the functions $f$ and $g$
respectively, $\xi(R_{GC}, t)$ can be re-written as:

\begin{equation}
\xi(R_{GC}, t)= a(R_{GC})\times \log(t) +b(R_{GC})
\end{equation}

We  fit straight  lines  to the  value  of $\xi(t)$  derived with  the
simulations between $\sim$3 and 10 Gyr.

 \begin{figure}
  \includegraphics[width=\columnwidth]{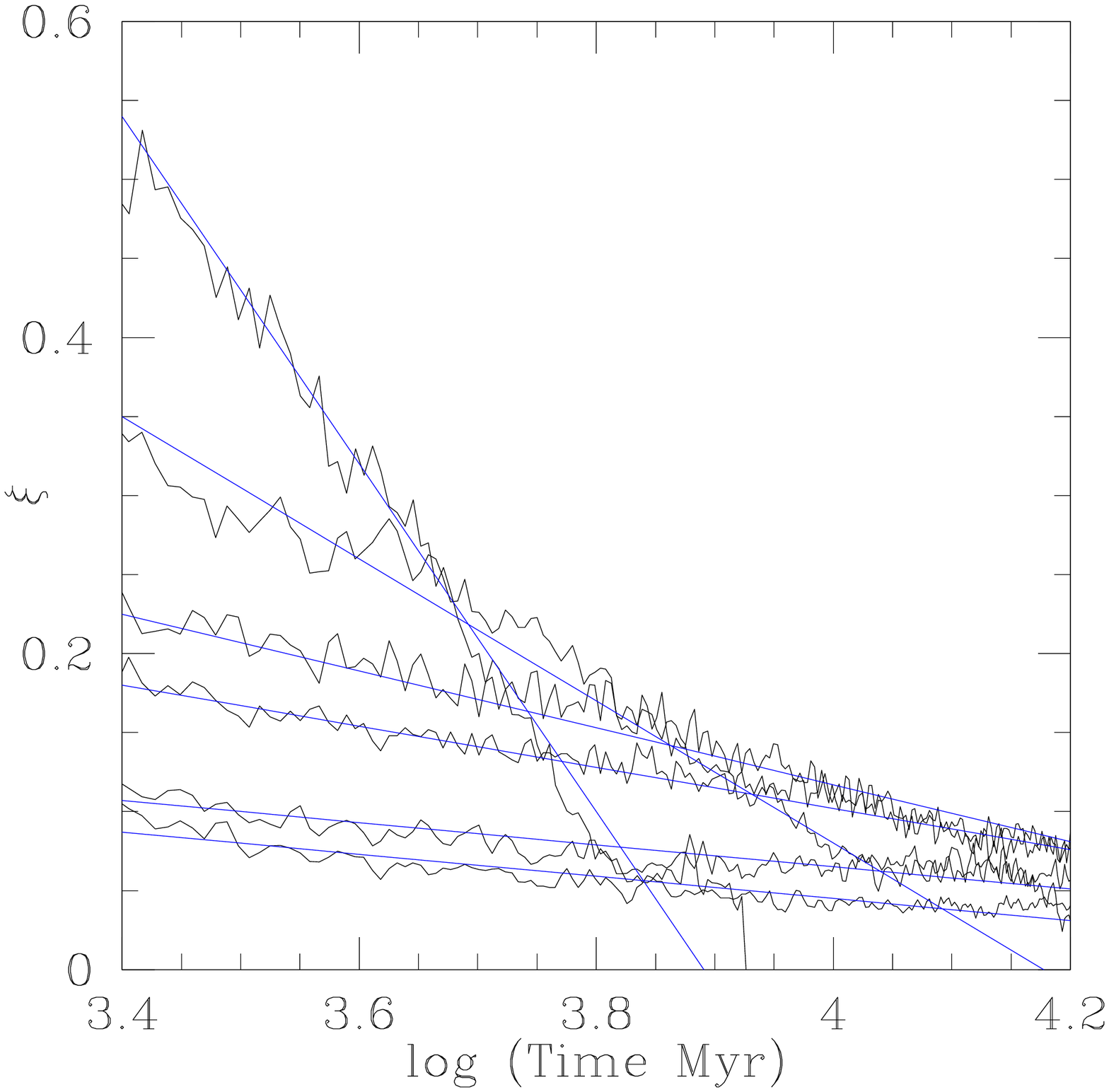}
 \caption{Fits to the values of $\xi$ for the following galactocentric 
distances, from top to bottom: 4, 6, 8, 10, 20, 50 kpc.
 \label{appendixfig1}}
 \end{figure}

 \begin{figure}
  \includegraphics[width=\columnwidth]{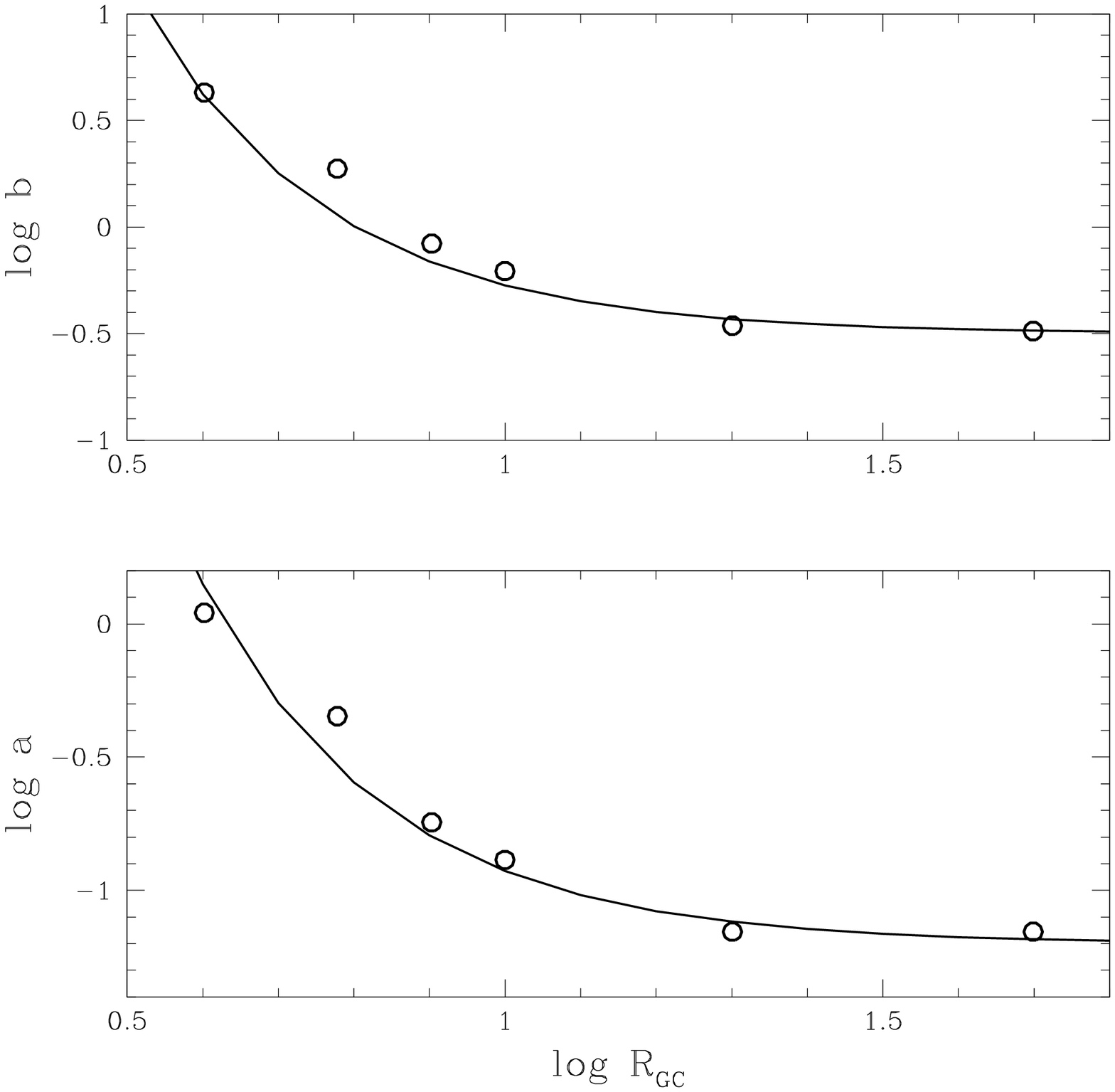}
 \caption{Slopes (i.e.\  $a$) and y intercepts  (i.e.\ $b$) of
the fitted lines (in blue in Figure 1 of the appendix). The solid lines show 
least-squares fits to the data in ln-ln-space.
 \label{appendixfig2}}
 \end{figure}

We then  take the slopes (i.e.\  $a$) and y intercepts  (i.e.\ $b$) of
the fitted lines (in blue in Figure 1 of the appendix). The slopes and
intercepts define six points (one for each galactocentric distance) in
the $\log (R_{GC})-\log(a)$  and $\log (R_{GC})-\log(b)$ space.  Using least-squares 
linear regression, these points are described by the following expressions::

\begin{equation}
a(R_{GC})=-10^{(3.0\times \exp(-4.0\times \log (R_{GC} -1.05)) -1.2)}
\end{equation}

and 

\begin{equation}
b(R_{GC})=10^{(2.5\times \exp(-4.0\times (\log (R_{GC} -0.4)) -0.5)}
\end{equation}

We thus can finally write $\xi(R_{GC},t)$ as

\begin{equation}
\xi(R_{GC},t)= a\times\log(t) + b
\end{equation}

with $t$ in Myr and $R_{GC}$ in kpc.

As  mentioned  above this  expression  of  $\xi(R_{GC},t)$ is  valid
between $\sim$3 and  10 Gyr. For a representation  of $\xi(R_{GC},t)$
between  0 to  3  Gyr we  suggest  a linear  interpolation between  an
initial global value of 6.9 to the  value given by Eq.\ 3 at 3 Gyr for
the orbit  of choice. A minimum  of 0.05 can  be applied to Eq.\  3 to
make it valid at later times.

The equations derived for  $\xi(R_{GC},t)$ are meant as an example of 
the proposed method for calculating $\xi(t)$ from a series of cluster simulations, for a 
given set of initial conditions.  It is valid for a given initial half-mass radius $r_{hm,i}$ 
(and concentration) and initial total cluster mass $M_i$.  We fix these two quantities in 
deriving our equations in this Appendix, which results in different initial degrees of tidal filling 
as a function of galactocentric distance.  This is meant to provide a reasonable, albeit 
approximate, method of calculating a solution to the dynamic mass loss rate (which changes 
over time due to various effects, such as stellar evolution) as $R_{GC}$ decreases due to 
dynamical friction, by iterating between the equations for $\xi(R_{GC},t)$ provided by our N-body 
simulations evolved at different $R_{GC}$.  Hence, our solution for $\xi(t)$ for a globular 
cluster with a given set of initial $r_{h,i}$ and $M_i$ values should appear as a line tilted
in the time-$\xi$ plane in the figure shown in this Appendix.  The slope of this line should change if the rate of 
decrease of galactocentric distance due to dynamical friction changes.   

In future work, we intend to populate a grid of $\xi(R_{GC},t)$ solutions, 
by obtaining the parameters  $a$ and $b$ in our fitting functions for different $r_{hm,i}$ and $M_i$
values in  additional N-body simulations.  That is, we will apply the same method as described 
in this paper to analogous sets of $N$-body models with different $r_{h,i}$ and $M_i$ values.
This will populate the parameter space for $\xi(t)$ relevant to the Milky Way globular cluster
population.  Our method folds into a single $\xi$ parameter all of the relevant physics (e.g., 
stellar evolution mass loss, relaxation-driven mass loss, etc.), such that a single analytic function describes the complex interplay between 
these physical processes. 

Populating a grid in $r_{hm,i}$ and $M_i$ with simulated clusters would be prohibitively 
expensive using $N$-body simulations, but entirely feasible using the Monte Carlo method for globular 
cluster evolution. Further justifying our comparison to the MOCCA simulations in 
Section 5.



\bsp	
\label{lastpage}
\end{document}